\documentclass[11pt]{article}

\usepackage{natbib}
\usepackage{fullpage, color, latexsym, amsmath,amssymb,amsthm,amsfonts, hyperref}

\newtheorem{theorem}{Theorem}[section]
\newtheorem{lemma}{Lemma}[section]
\newtheorem{corollary}{Corollary}[section]

\newtheorem{definition}{Definition}[section]

\newtheorem{assumption}{Assumption}[section]

\newcommand{\R}{{\mathbb{R}}}
\newcommand{\Expct}{{\mathbb E}}

\newcommand{\calE}{{\cal E}}

\newcommand{\calM}{{\cal M}}

\newcommand{\calP}{{\cal P}}
\newcommand{\calQ}{{\cal Q}}
\newcommand{\calR}{{\cal R}}

\newcommand{\calT}{{\cal T}}

\newcommand{\sumt}{{\sum_{t=1}^{\infty}}}

\newcommand{\sumi}{{\sum_i}}
\newcommand{\sumk}{{\sum_{k=0}^{\infty}}}

\newcommand{\rinit}{r}

\newcommand{\experience}{{e}}
\newcommand{\experiencet}{e_{t}}
\newcommand{\experiencei}{e_{i}}
\newcommand{\experienceit}{e_{i,t}}

\newcommand{\experienceRt}{\hat{e}_{t}}
\newcommand{\experienceRit}{\hat{e}_{i,t}}

\newcommand{\Report}{{R}}
\newcommand{\Reporti}{{R_i}}

\newcommand{\Info}{{\cal P}}
\newcommand{\info}{{\rho}}
\newcommand{\infoit}{{\rho_{i,t}}}
\newcommand{\infot}{{\rho_{t}}}
\newcommand{\infoi}{{\rho_{i}}}

\newcommand{\infotplus}{{\rho_{t+1}}}

\newcommand{\allocit}{q_{i,t}}

\newcommand{\galloc}{{q^{z}}}

\newcommand{\gallocit}{{q_{i,t}^{z}}}

\newcommand{\gValuei}{{V_i}}
\newcommand{\gPricei}{{P_i}}

\newcommand{\valuei}{{v_i}}

\newcommand{\virtual}{{\psi}}
\newcommand{\virtuali}{{\psi_i}}

\newcommand{\statet}{\experiencet,\infot}

\def\thetaR{\hat{\theta}}
\def\thetaRz{\hat{\theta}_0}
\def\thetaRit{\hat{\theta}_{i,t}}
\def\thetaRiz{\hat{\theta}_{i,0}}
\def\thetaRi{\hat{\theta}_{i}}
\def\thetaRnoti{\hat{\theta}_{-i}}
\def\thetaRt{\hat{\theta}_{t}}

\def\rev{\mbox{Rev}}

\newcommand{\deltat}{\delta^{t-1}}

\def\GI{{\cal G}}

\def\i{\hspace*{7mm}}

\title{
An Optimal Dynamic Mechanism\\ for Multi-Armed Bandit Processes
}

\author{
Sham M. Kakade\thanks{Department of Statistics, Wharton School, University of Pennsylvania -- skakade@wharton.upenn.edu.
Work performed in part at Microsoft Research.}
\and Ilan Lobel\thanks{Stern School of Business, New York University -- ilobel@stern.nyu.edu. Work performed in part at Microsoft Research.}
\and Hamid Nazerzadeh\thanks{Microsoft Research, New England Lab -- hamidnz@microsoft.com}
\footnote{
We would like to thank Maher Said for many fruitful discussions.
}
}

\begin{document}
\maketitle

\begin{abstract}
  We consider the problem of revenue-optimal dynamic mechanism design
  in settings where agents' types evolve over time as
  a function of their (both public and private) experience with items that are auctioned repeatedly over an infinite horizon.
  A central question here is understanding
  what natural restrictions on the environment permit the design of
  optimal mechanisms (note that even in the simpler static setting,
  optimal mechanisms are characterized only under certain restrictions).
  We provide a {\em structural characterization} of a natural
  ``separable'' multi-armed bandit environment (where the evolution
  and incentive structure of the a-priori type is decoupled from the
  subsequent experience in a precise sense) where dynamic optimal mechanism design is possible. Here, we present the
  Virtual Index Mechanism, an optimal dynamic mechanism, which
  maximizes the (long term) {\em virtual surplus} using the classical
  Gittins algorithm. The mechanism optimally balances exploration and
  exploitation, taking incentives into account.

  We pay close attention to the applicability of our results to the (repeated) ad auctions used in sponsored search,
  where a given ad space is repeatedly allocated to advertisers. The value of an ad allocation to a given advertiser depends on multiple factors such as the probability that a user clicks on the ad, the likelihood that the user performs a valuable transaction (such as a purchase) on the advertiser's website and, ultimately, the value of that transaction. Furthermore, some of the private information is learned over time, for example, as the advertiser obtains better estimates of the likelihood of a transaction occurring. We provide a dynamic mechanism that extracts the maximum feasible revenue given the constraints imposed by the need to repeatedly elicit information.

 One interesting implication of our results is a certain
  revenue equivalence between public and private experience, in these
  separable environments. The optimal revenue is no less than if agents'
  private experience (which they are free to misreport, if they are
  not incentivized appropriately) were instead publicly observed by the mechanism.
\end{abstract}

\newpage

\section{Introduction}

Designing mechanisms in \emph{dynamic} environments --- in which agents
valuations evolve as a function of their ``experience'' with the
allocated items
--- is a problem which has received much recent
interest. One of the most compelling applications here is that of ad auction
sponsored search,
in which search engines sell the advertisement spaces  that appear alongside the search results.

Let us discuss the sponsored search example in more detail:
typically, an advertiser places an ad in order to: first, draw a
client to visit the advertiser's website (via a click on the displayed
ad), and then, subsequently, have the client purchase some
product.  The expected value that an advertiser obtains from a
displayed ad depends on \emph{both} the ``click-through rate'' (the
probability that a user clicks on the ad, sending the user to the
advertiser's website) and the ``conversion rate'' (the probability
that the user who visits the website performs a desired transaction,
e.g. a purchase).  This is a dynamic environment in which both
advertisers and the search engine (the mechanism) learn and update their
estimates of these rates over time.  Observe that a click is a {\em
  public experience}, i.e., observed by both the advertiser and the
search engine.
In contrast, a transaction is only observed by the advertiser --- it's
{\em private experience} of the advertiser with the displayed ad. The dynamic challenge here is to design appropriate
mechanisms which align incentives such that the search engine and the
advertiser share this information for some desired outcome.

In the oft used practical mechanisms, the learning of
click-through-rates and conversions-rates have been separated due to
this asymmetry of information --- (see \cite{MahdianT07,AgrawalAY09}
for further discussions on ``pay-per-action" pricing schemes). Two
fundamental question that arises in this setting is: how much revenue
does this asymmetry of information cost the mechanism?
How much more revenue would the mechanism be able to obtain,
if it were able to monitor the transactions on the advertisers'
websites?

In the static setting, the two foremost objectives for a mechanism are
either maximizing the social welfare of the buyers (\emph{efficiency}) or the
maximizing the revenue of the seller (\emph{optimality}) --- though the
spectrum of other objectives is large and notable. By extension, these
are the natural two objectives to consider for the dynamic setting.
With regards to maximizing the future social welfare in a dynamic
setting, there is an elegant extension of the efficient (VCG) mechanism applicable
to quite general dynamic settings by \cite{ParkesS03, BergemannV07}.
This dynamic mechanism seamlessly inherits the core
concepts of the static VCG mechanism --- namely, charging an agent the
externality they impose, which is implemented via dynamic programming
ideas.  Related dynamic mechanisms include the dynamic budget-balanced
efficient mechanism by~\cite{AtheyS07}, efficient mechanisms for
dynamic populations by~\cite{CavalloPS07}, and non-Bayesian
(asymptotically) efficient dynamic
mechanisms~(see \cite{NazerzadehSV08,BabaioffSS09}).

With regards to \emph{optimal} dynamic mechanisms in a dynamic
setting, the state of affairs is more murky. As we discuss in the next
section, while there are detailed characterizations of necessary
conditions for which (incentive compatible) dynamic mechanisms must
satisfy, there are only a few rather restricted special cases for
which optimal mechanisms are characterized (e.g. see
\cite{PavanST08}).  To some extent, results for special cases are
to be expected, as even in the simpler static mechanism design problem, the efficient
(VCG) mechanism is applicable to general settings (e.g. combinatorial auctions
with no distributional assumptions) while even the optimal mechanism
for selling a single item (provided in the seminal work of \cite{Myerson81})
 is only applicable under certain distributional restrictions

In the more challenging dynamic setting, perhaps the most central
question is understanding what natural restrictions permit the design
of optimal mechanisms. This is the focus of this paper, and we provide
a certain structural characterization of a ``separable'' environment (allowing
for public and private experience), in which optimal dynamic mechanism design is possible.
Our characterization is rather rich in that it permits both a natural
stochastic processes (where private and public signals can be discrete or from
abstract spaces) and is applicable to certain natural formalizations of the
aforementioned sponsored search setting (where both
public click-through-rates and private purchase-rates evolve over
time).  Our construction draws a rather close connection
to efficient mechanism design (where our optimal mechanism utilizes
the efficient mechanism for a certain affinely transformed social
welfare function). Furthermore, we also address
the issue of how much revenue is lost due to private signals (rather than
public signals, observed by the mechanism) in these separable
environments -- somewhat surprisingly, there is no loss.

\subsection{Contributions}
Our main contribution is designing an individually rational and incentive compatible, revenue-optimal mechanism,
called the {Virtual Index} Mechanism,
for settings with $1$ seller and $k$ (agents) buyers where the environment
 satisfies certain \emph{separable} properties and
evolves according to a \emph{multi-armed bandit process}.

The Virtual Index Mechanism is quite simple.
In short, the allocation rule of the mechanism is based on the notion
of Gittins indices (see~\cite{Gittins89, Whittle82})
and the payment rule is derived by considering a dynamic VCG mechanism (where the
social welfare function is transformed under a particular,
time-varying affine function).

The allocation assigns to each agent an ``index'' which is computed based on solely
 the agent's current state;
and at each step the mechanism allocates the item to an agent with the
highest index. As in~\cite{Gittins89, Whittle82}, the key observation
is that this computation does not require specifying a policy in terms of
the (potentially exponentially many) histories.
If all the agents are truthful, then the mechanism maximizes the ``virtual surplus";
the idea pioneered in \cite{Myerson81}.

It turns out that the allocation we use also coincides with the efficient
dynamic (VCG) mechanism (with respect to a transformed social welfare
function). Due to this, the payment rule of our mechanism is rather
simple to specify. In fact, one of our technical contributions is using this
reduction to dynamic (affine) VCG in the construction of a revenue-optimal mechanism. This
connection is useful for two reasons: first, it allows us to reduce
the problem of checking incentive compatibility to essentially a one
period problem. Second, this VCG pricing allows us to utilize rather
general multi-armed bandit processes.

A surprising implication of our result is that the seller does not
lose any revenue (under the optimal mechanism) if the experience were
private rather than publicly observed.
In the context of sponsored search,
this implies that (in environments which satisfy our ``separability" assumption)
the ability to monitor the transactions that occur
in the advertisers' own websites does not increase the revenue.
An important business insight provided by this result is that pay-per-action mechanisms can be
implemented without a loss of revenue if the search engine is able to
commit to a long-term contract.

\subsection{Related Work}
The most closely related work to ours is that in \cite{PavanST08,PavanST09}.~\footnote{We note that the work in
  ~\cite{PavanST09} is a recent preliminary draft, and is concurrent to this
  work.}
The primary contributions of this work is that it
establishes rather detailed \emph{necessary} conditions for dynamic
incentive compatibility in both finite horizons~(\cite{PavanST08}) and
infinite horizons~(\cite{PavanST09}).
In addition, \cite{PavanST08} also establish a dynamic version of the Revenue
Equivalence Theorem in a particular finite horizon setting.
With regards to providing dynamic optimal mechanisms, their work only
provides mechanisms in somewhat limited special cases --- such as when
valuations evolve according to a certain auto-regressive $AR(k)$ stochastic processes,
and, in \cite{PavanST09}, the value evolution evolves according to a
particular additive manner, where each private experience of the
agents is assumed to be independent of all previous private
experiences (it is allowed to depend only on the number of times the
item was previously allocated, a much more restrictive assumption than
those provided by our results).  We should also note the work in
\cite{Deb08}, which provides an optimal mechanism in a restricted
setting where the value is Markov in the previous value, among other
technical conditions --- again, their model does not permit rich
dependencies on historical signals.  In our sponsored
search example, these prior results are not applicable, due to both the
multiplicative nature of the value function (as discussed later), and
due to that the sequence of experiences are not independent (e.g. with
a Bayesian, ``Bernoulli'' prior on the probabilities of binary ``click'' or
``purchase'' events, the experiences are not necessarily independent).

Also, in contrast to \cite{PavanST08,PavanST09}, we should emphasize that the aim of
our work is \emph{not} to characterize necessary conditions which any
(incentive compatible) dynamic mechanism must satisfy -- our focus is on
the optimal mechanism itself. In fact, we do
not even utilize the dynamic ``envelope'' conditions provided by
\cite{PavanST08,PavanST09}, as they require: many detailed technical
assumptions; the signals (the experiences) to be real valued;
and, often, certain probability kernels to have densities and be differentiable --- as
such, these conditions either do not hold or are difficult to verify in our
setting. Instead, our derivation proceeds from merely static
considerations, where we use only incentive compatibility constraints
from static mechanism design theory (see \cite{MilgromS02}) to
establish the expected revenue of any (incentive compatible) mechanism
(e.g. the so called ``envelope theorem'').  Certainly, the sufficient
conditions provided by~\cite{PavanST08,PavanST09} (derived with rather
sophisticated proofs) are stronger than than those conditions used
here, as they explicitly account for dynamic considerations and are
interesting in their own right --- one further direction is if these
conditions can be used to derive optimal dynamic mechanisms in
settings more general than those provided here.

Conceptually, our proof is rather simple: both the connection to
dynamic (affine) VCG and our use of only a static ``envelope''
condition allow us to reduce the proof of dynamic incentive
compatibility to essentially a one-period, static problem. However,
this one-period verification requires a delicate stochastic coupling
argument, where we utilize both the bandit nature and the separability
of our stochastic process.

We also briefly mention other notable work here. \cite{VulcanoRM02}
analyze the problem of optimal dynamic mechanism design in the context
of perishable goods (work later expanded on by \cite{PaiV08}). These
works built upon dynamic programming ideas to extend the classical
result of \cite{Myerson81} to a dynamic setting.  A key assumption in
both of these papers is that agents' valuations do not evolve over
time.  \cite{Battaglini05} studies the question of optimal mechanism
design in a setting with a single consumer whose private information
is given by a 2-state Markov Chain.  \cite{EsoS07} obtain a result in a
two-period model that is similar in flavor to ours: the buyers do not benefit from
obtaining new private information at period 2 if the seller uses a `handicap' auction, 
where the handicap is given by the buyer's virtual value at period 1.

\subsection{Organization}

We organize our paper as follows. In Section \ref{sec:model}, we
formalize our model, define separable environments, show examples and
define the basic notions we use throughout the paper, such as
incentive compatibility and optimality of mechanisms. In Section
\ref{sec:incentives}, we consider a variant of our model where the
mechanism can monitor the private experiences of the agents and
describe how to it establishes a bound for the revenue of a mechanism
in our setting. In Section \ref{sec:mechanism}, we state and explain
our main theorem --- the optimality of the Virtual Index Mechanism. All proofs are in the Appendix.

\section{Preliminaries} \label{sec:model}

\subsection{Environment}

We consider a setting with $1$ seller and $k$ agents (buyers) who are
competing for items that are being allocated at every timestep
(starting at $t=1$) over a discrete time
infinite horizon.  At the start of $t=1$, agents (privately) learn their initial
types. The {\em initial type} of agent $i$ is a (non-negative) real
number $\theta_i\in [0,\overline{\Theta}_i]$, independently distributed according to some
given distribution $F_i(\cdot)$.  At every subsequent timestep, the
state of each agent $i$ is summarized by the tuple of their initial
type $\theta_i$ and their (subsequent) ``experience'' with the item --- this
experience summarizes the type of the agent due to interactions with
the item and the experience need not be real valued.  More precisely, agent $i$'s state at time $t$ is of the
form $(\theta_i,\experienceit,\infoit)$, where the current state of
the private experience is denoted by $\experienceit\in \calE_i$ and
the public experience is denoted by $\infoit\in\Info_i$, where $\calE_i$
and $\Info_i$ are some (potentially \emph{arbitrary}) set. Here, only
the agent observes their private experience, while the public experience is
also observed by the mechanism.

We should emphasize that the first type $\theta$ is
real for reasons similar to that in the static setting --- derivations of
optimal mechanisms typically involve calculus on real valued
types. However, it is \emph{only} this first type that we assume to be
real valued (subsequent experience is allowed to live in arbitrary
signal spaces). As we specify later, this first type $\theta$ has a
persistent effect on the incentives (e.g., the values after $t=1$ could
also depend on the initial type).

If agent $i$ is allocated, the state of agent $i$'s public and private
experience with the item evolve in a Markovian manner. If $i$ is not
allocated, then the experience {\em does not change} --- in this
sense, we are dealing with a Markovian ``bandit'' process. By nature
of public information, we assume the public process is completely
decoupled from private information, e.g., the probability that the
next public experience is $\info_i'$ conditioned on the current
experience being $\info_{i}$ is $G(\info_i'|\info_{i})$. In the most
general sense, the evolution of the private experience could depend
on the entire current state, e.g., the probability that the next
private experience is $\experience_i'$ conditioned on the current
state $(\theta_i,\experience_i,\info_i)$ is
$H(\experience_i'|\theta_i,\experience_i,\info_i)$. However, it turns
out that are not able to handle this level of generality, and our
structural characterization specifies certain natural restrictions (in
the next subsection).

Note that the public experience evolution process
only depends on the public times series, while the private experience
process is allowed to depend on both private and public experience. We assume that private experience is only
observed by the agent, but the public information is also observed by
the mechanism. At time $t=1$ (prior to the allocation at $t=1$), for every agent $i$,
 $\experience_{i1}=\emptyset$ is the empty experience, known by both the agent and the mechanism.

The (instantaneous) value of each agent $i$ at time $t$ is a
(stationary) function of their current state. In particular, agent
$i$'s value is $\valuei(\theta_i,\experienceit,\infoit)$ when in state
$(\theta_i,\experienceit,\infoit)$ at time $t$. The expected (future)
value of agent $i$ for the item at time $t$ is equal to $\deltat
\valuei(\theta_i,\experienceit,\infoit)$ where $\delta$, $0<\delta<1$,
is the common discount factor.

\subsection{Separable Environments} \label{sec:separable}

Characterizing the assumptions which permit optimal mechanisms design
is perhaps the most central question --- even in the static setting, only in
(natural) special cases optimal mechanisms are known. In our dynamic setting, it turns
that we are not able to characterize an optimal mechanism in the
full generality of the above ``bandit'' environment. However, our main contribution
is specifying a natural ``separable'' environment, under which we can
derive an optimal mechanism. We say that the environment is
\emph{separable} if \emph{both} the stochastic process over the types
and the value functions themselves are separable, in a precise sense
which we now define. Intuitively, the notion of separability decouples
the initial (real valued) type $\theta$ from the experience, both in terms of the
stochastic evolution and the incentive structure.

\begin{definition}
The stochastic process is said to be separable if the evolution of the
private experience is Markovian in the current experience,
e.g., the probability that the next
private experiences is $\experience_i'$ conditioned on the current
state $(\theta_i,\experience_i,\info_i)$ is
$H(\experience_i'|\experience_i,\info_i)$ (in particular, $H$ does not
depend on $\theta_i$).
\end{definition}

We consider two natural classes of separable value functions.

\begin{definition}  \label{def:separable}
Additively or multiplicatively {\em separable} value functions are
defined as follows:
\begin{itemize}
\item
An \emph{additively separable value} function has the following functional form, for all $i$, $\theta_i$, $\experience_i$, and $\info_i$: \\
\[\valuei(\theta_i,\experience_i,\info_i) = A_i(\theta_i,\info_i)+B_i(\experience_i,\info_i)\]
\item A \emph{multiplicatively separable value} is of the form: \\
\[\valuei(\theta_i,\experience_i,\info_i) = A_i(\theta_i)B_i(\experience_i,\info_i) -C_i(\info_i)\]
\end{itemize}
\end{definition}

Taken together, we say that the environment is (additively or multiplicatively) separable.

\subsection{Examples of Separable Environments}

We now provide two examples of settings for separable value
functions, which fall within our framework (and satisfy our assumptions).

\medskip

{\bf Sponsored Search:}
Consider an auction for a keyword that corresponds to a certain product.
Suppose $i$ is an online retailer of such a product who participates in the corresponding sponsored search auction. Every time a user types in the keyword, the ad space is allocated to (at most) one retailer.
Every time a user purchases the product from them, the retailer $i$ obtains a value of $\theta_i$ (and $0$ otherwise).
The private experience $\experienceit$ describes the retailer's Bayesian belief about the probability of a purchase given a click has occurred.
Similarly, the public experience $\infoit$ represents the Bayesian belief about the probability of a click occurring given the retailer's ad is shown.
Therefore, $\valuei(\theta_i,\experienceit,\infoit) = \theta_i \Pr[\mbox{purchase}|\experienceit, click] \Pr[\mbox{click}|\infoit]$.
After each time the ad of retailer $i$ is shown to a user, both the retailer and the search engine update the belief $\infoit$ about probability of a click. After each click, the retailer updates their belief  $\experienceit$ about the probability  of a purchase.

\medskip
{\bf Auto-Regressive (AR):}
The evolution of the valuation of each agent $i$ in an $AR(1)$ model is as follows.
The initial value of agent $i$ is given by $v_{i,0} = \theta_i$, and every time the item is allocated to agent $i$ his valuation is updated according to $v_{i,t+1} = A_i v_{i,t} + B_i(e_{i,t}, \infoit)$.
In the $AR$ model considered in \cite{PavanST08, PavanST09}, the agent's value is updated by adding an independent shock, or a shock that depends only on the previous allocations (e.g., number of times the item was allocated to the agent), but is independent of all private information. Our model allows for the update of the value to depend both on the private and the public experiences. They also consider $AR(k)$ processes, where the valuation is updated according to an affine function of the values at the $k$ previous times the item was allocated to the agent. Our restriction to $AR(1)$ models is without loss of generality since, by augmenting the state space, an $AR(k)$
process can be represented as an $AR(1)$ process (and through an
appropriate choice of the function $A_i(\cdot)$).

\subsection{Mechanisms, Incentive Constraints, and Optimality}
By the Revelation Principle~(cf.\ \cite{Myerson86}), without loss of generality we can focus on direct mechanisms.\footnote{
The Revelation Principle implies that an equilibrium outcome in any
indirect mechanism can also be induced as an equilibrium outcome of an (incentive compatible) direct mechanism.
}
A direct mechanism $\calM(\calQ, \calP)$ is defined by a pair of an allocation rule $\calQ$ and a payment rule $\calP$.
In a dynamic direct mechanism, at each timestep $t$,
an agent is asked to {\em report} their current private state pair
$(\theta_i,\experienceit)$ --- we denote this report by $(\thetaRit,\experienceRit)$.
Note that a direct mechanism elicits redundant information as the
initial type $\theta_i$ of an agent remains constant over time, while
the mechanism asks the agent to re-report this type every round
(similarly, the private experience of an agent does not evolve in a
period in which it did not receive an allocation, yet the mechanism asks for re-reports).

We denote the (joint) vector of reports, public state, allocations, and payments
at time $t$ by $(\thetaRt, \experienceRt, \info_t, q_t, p_t)$, where
$q_{i,t}$ and $p_{i,t}$ correspond to the allocation and payment of
agent $i$ at time $t$ ($q_{i,t}=1$ if $i$ received the item at time $t$
and $0$ otherwise).
The history $h_t$ observed by the seller at any given time $t$ includes the past
reports, the past public experience, the past allocations, and the
past payments (and does not include either the past private experiences or
the initial types). The
history $h_{i,t}$ observed by an agent $i$ at time time $t$ includes her past initial type, her
private experiences, her prior reports, the prior payments, and the
prior allocations (and does not include other agents true or reported
private experiences or initial types).

In short, at each timestep $t\ge 1$ the following sequence of events
occur:
\begin{enumerate}
\item Each agent $i$ reports $(\thetaRit, \experienceRit)$ only to the mechanism.
\item The mechanism allocates the item to an agent $i^\star$, if $q_{i^\star,t}= 1$ (or potentially to no one).
\item Each agent $i$ is charged $p_{i,t}$.
\item Agent $i^\star$'s (private and public) experience evolve. 
\end{enumerate}

We now define the incentive constraints of the mechanism.  First, some
definitions are in order. A reporting strategy for agent $i$ is a
mapping from her type, her private experience state, and the history
to a report (of her initial type and the current state of the private
experience).  Let $\Report$ denote a joint reporting strategy and
$\Report_i$ denote this strategy for $i$. For mechanism $\calM$,
define the (discounted) expected future value and payment of agent $i$
at time $t$, under (joint) reporting strategy $\Report$,
conditioned on some event $h$, as follows:
\begin{eqnarray*}
V_{i,t}^{\calM,\Report}(h) =
\Expct\left[ \sum_{t'=t}^\infty \delta^{t'-1} q_{i,t'} \valuei(\theta_i, e_{i,t'},\rho_{i,t'}) \middle\vert h\right]\\
\end{eqnarray*}
\begin{eqnarray*}
P_{i,t}^{\calM,\Report} (h) =
\Expct\left[\sum_{t'=t}^\infty \delta^{t'-1} p_{i,t'} \middle\vert h \right]
\end{eqnarray*}
where the evolution of the process is under $\calM$ under reporting
strategy $\calR$ --- conditioned on the event $h_t$ (the expectation
is with respect to all variables not conditioned on). For example, for
some current state $\theta_i,\experienceit,\infoit$ for $i$ and
history $h_{i,t}$ for agent $i$,
$V_{i,t}^{\calM,\Report}(\theta_i,\experienceit,\infoit,h_{i,t})$ is
the expected future value of $i$ conditioned on her knowledge at time $t$.
Similarly, define the
(discounted) {\em expected future utility} for agent $i$ as:
\begin{equation}    \label{eq:utility}
U_{i,t}^{\calM,\Report} (h)
= V_{i,t}^{\calM,\Report}(h) -  P_{i,t}^{\calM,\Report} (h)
\end{equation}

We say that $R_i$ is a best response to $R_{-i}$ conditioned on event $h$
(for agent $i$) if the $R_i$ maximizes her utility,
e.g., $U_{i,t}^{\calM,\Reporti} (h) $ is
greater than $U_{i,t}^{\calM,\Reporti'}(h) $ (for all other $\Reporti'$, where $\Report_{-i}$ is held
fixed).  We say the {\em truthtelling strategy} $\calT$ is the
reporting strategy under which all agents always reports their initial
types and their private experiences truthfully.

We now define incentive compatibility. Roughly speaking, this concept
says that as long as all agents are truthful, then no agent ever wants to deviate.
\begin{definition} (Incentive Compatibility)
A dynamic direct mechanism is {\em incentive compatible} if, for each agent $i$, with
probability one, truthtelling
is a best response (assuming the other agents to be truthful) at each
time $t$ with respect to the history of $i$ at time $t$. Precisely,
with probability $1$,  for all times $t$ and all $\Reporti$,
\[
U_{i,t}^{\calM,\calT}(\theta_i,\experienceit,\infoit,h_{i,t}) \geq
U_{i,t}^{\calM,(\Reporti,\calT_{-i})}(\theta_i,\experienceit,\infoit,h_{i,t})
\]
where the probability is with respect to
$\theta_i,\experienceit,\infoit,h_{i,t} $ sampled under the truthful
reporting strategy.
\end{definition}

We consider a stronger notion, namely, periodic ex-post incentive
compatibility, where best responses hold even on histories where
misreports occur (see Definition~\ref{def:XIC}).

We also allow the following participation constraint, in which
agents may opt out at any time for $0$ future utility.

\begin{definition} (Individual Rationality)
Under an {\em individually rational} mechanism, for each agent $i$, with probability 1, truthful agents obtain
a non-negative expected future utility  assuming the other agents are truthful. Precisely,  with probability $1$,  for all times $t$ and all $\Reporti$, \[U_{i,t}^{\calM,\calT}(\theta_i,\experienceit,\infoit,h_{i,t}) \geq 0\]
where the probability is with respect to
$\theta_i,\experienceit,\infoit,h_{i,t} $ sampled under the truthful
reporting strategy.
\end{definition}

The {\em expected revenue} of an incentive compatible mechanism under the
truthful strategy is the discounted sum of all payments of the agents, i.e.,
\begin{equation}    \label{eq:revenue}
\rev^{\calM} = \Expct\left[ \sum_{i=1}^k P_{i,1}^{\calM,\calT}(\theta)\right]
\end{equation}
The objective of the seller is to maximize this expected revenue,
subject to both the incentive compatibility constraint and rationality
constraint.  Precisely,
\begin{definition} (Optimality)
An individually rational and incentive compatible mechanism is {\em optimal} if it maximizes the expected revenue among all individually rational and incentive compatible mechanisms.
\end{definition}

\section{A Dynamic Constraint From Static Considerations}\label{sec:incentives}
In this section, we establish a bound on the maximum revenue a seller can obtain given the incentive constraints of the agents. We consider a modified setting that we call the complete dynamic monitoring problem and show that it determines a bound for the seller's revenue in the model defined in Section \ref{sec:model}.

Consider the modified setting where the seller can fully monitor the agents' private experiences $\{\experience_{i,t}\}$, for each agent $i$ and all $t > 0$. The seller still cannot observe the initial types $\theta_i$'s, and, as before, agents report $\hat{\theta}_i$ to the mechanism in the initial period. We denote this setting the \textbf{complete dynamic monitoring} problem. In this new setting, the mechanism design problem is a completely static one. Therefore, the incentive compatibility results of \cite{MilgromS02} for static mechanism design apply here. The following ``revenue equivalency" theorem establishes the revenue obtained by an incentive compatible mechanism in the complete dynamic monitoring setting; the proof is in Appendix~\ref{sec:envelope}.

\begin{theorem} \label{thm:ICmonitoring}  Assume complete dynamic monitoring. Assume as well that the the partial derivative $\frac{\partial  \valuei(\theta_i,\experiencei,\infoi)}{\partial \theta_i}$ exists for all  $\theta_i$, $\experiencei$ and $\infoi$ and there exists some $B < \infty$ such that $|\frac{\partial  \valuei(\theta_i,\experiencei,\infoi)}{\partial \theta_i}| \leq B$ for all  $\theta_i$, $\experiencei$ and $\infoi$. Then, the revenue $\rev^\calM$ of any such incentive compatible mechanism $\calM$ satisfies
\begin{equation} \label{eq:static IC}
 \rev^\calM = \sum_{i=1}^k \Expct\left[\sumt \deltat \allocit \virtuali(\theta_i,\experienceit,\infoit) - U_i^\calM(0,\theta_{-i})\right]
\end{equation}  where $\virtuali$ is defined as
\begin{equation} \label{eq:virtual_value}
\virtuali(\theta_i,\experienceit,\infoit) = v_i(\theta_{i}, \experienceit,\infoit)
- \frac{1-F(\theta_i)}{f(\theta_i)} \frac{\partial v_i(\theta_i, \experienceit,\infoit)}{\partial \theta_i}
\end{equation} and $U_i^\calM(0,\theta_{-i})$ is the utility agent $i$ obtains if his type is equal to 0 and the other agents' types are $\theta_{-i}$.
\end{theorem}

Similarly to \cite{Myerson81}, we refer to $\virtuali$ as the {\em virtual value}. The right-hand side of Eq.\ (\ref{eq:static IC}) is  the {\em virtual surplus}. Individual rationality is equivalent to the requirement that $U_i^\calM(0,\theta_{-i}) \geq 0$ for all $i$ and $\theta_{-i}$. Therefore, Eq.\ (\ref{eq:static IC}) implies that for any incentive compatible and individually rational mechanism $\calM$ in the complete dynamic monitoring setting,
\begin{equation}\label{eq:rev monitoring}
 \rev^\calM \leq \sum_{i=1}^k \Expct\left[\sumt \deltat \allocit \virtuali(\theta_i,\experienceit,\infoit)\right]
\end{equation} if the assumptions of Theorem \ref{thm:ICmonitoring} hold.

 Now consider a direct mechanism $\calM$ for the original setting without monitoring which is both incentive compatible and individually rational. The exact same mechanism, including the allocation and payment rules, can be applied in the setting with complete dynamic monitoring. To do so, we simply replace the agent's reported private experiences $\{\hat{e}_{i,t}\}$ by the agent's (monitored) private experiences $\{e_{i,t}\}$ in the input of the allocation and payment rules. Any strategy available to the agents in the setting with complete dynamic monitoring is a feasible strategy in the setting without monitoring (where the agent reports truthfully after the initial report). Therefore, if all other agents are truthful, any profitable deviation from the truthful strategy in the setting with complete dynamic monitoring implies a profitable deviation in the setting without monitoring. Since no such profitable deviations exist in the setting without monitoring, we obtain that the mechanism $\calM$ is both incentive compatible and individually rational in the setting with complete dynamic monitoring. Therefore,
 Eq.\ (\ref{eq:rev monitoring}) establishes an upper bound on the revenue of mechanisms for the setting without monitoring as well.

\begin{corollary} \label{thm:upperbound} Assume the the partial derivative $\frac{\partial  \valuei(\theta_i,\experiencei,\infoi)}{\partial \theta_i}$ exists for all  $\theta_i$, $\experiencei$ and $\infoi$ and there exists some $B < \infty$ such that $|\frac{\partial  \valuei(\theta_i,\experiencei,\infoi)}{\partial \theta_i}| \leq B$ for all  $\theta_i$, $\experiencei$ and $\infoi$. Then, the revenue $\rev^\calM$ of any incentive compatible, individually rational mechanism $\calM$ satisfies
\begin{equation}\label{eq:virtual_surplus}
 \rev^\calM \leq \max_{\calQ \in \mathbb{Q}} \Expct\left[\sumt \sum_{i=1}^k \deltat \allocit \virtuali(\theta_i,\experienceit,\infoit)\right],
\end{equation} where  $\mathbb{Q}$ represents the set of all allocation rules.
\end{corollary}

The corollary suggests a candidate allocation rule for an optimal mechanism in the setting without monitoring.
The maximization problem on the right-hand side of Eq.~(\ref{eq:virtual_surplus}) is a multi-armed bandit problem, where the payoff of arms given by the virtual values. This optimization problem can be solved using Gittins indices~(see \cite{Gittins89,Whittle82}). We use this allocation rule in the mechanism we design in the next session.

\section{The Virtual Index Mechanism}\label{sec:mechanism}
In this section, we present our main result, an optimal dynamic mechanism, called the \emph{Virtual Index} Mechanism.
In short, the allocation rule is as follows: the mechanism assigns to
each agent an ``index" (computed based on virtual values) and at each
step allocates the item to an agent with the highest index.  If all
the agents are truthful then the mechanism maximizes the revenue as
well as the virtual surplus.  Furthermore, the mechanism enjoys more
desirable incentive constraints --- it satisfies stronger notions of
incentive compatibility and individual rationality.
\begin{definition} \label{def:XIC} (Periodic Ex-post Incentive Compatibility)
A dynamic direct mechanism is {\em  periodic ex-post incentive compatible} if
for all agents, truth-telling is a best response conditioned on
\emph{any} historical event and conditioned on the current state of
the other agents (assuming other agents to be truthful in the
future). Note here the historical event need not be a truthful history,
but is arbitrary.
\end{definition}
\begin{definition} (Periodic Ex-post Individually Rationality)
A dynamic direct mechanism is {\em  periodic ex-post individually
  rational} if for all agents and conditioned on
\emph{any} historical event and conditioned on the current state of
the other agents, the agent's expected future utility is non-negative under the
truthful strategy (assuming other agents to be truthful in the future).
\end{definition}

These two stronger incentive constraints are ensured by the
dynamic VCG mechanism provided by \cite{BergemannV07}. In our
setting, our optimal mechanism also enjoys these properties.

Our main theorem relies on the following:
\begin{assumption}\label{ass:seperable} (Separable Environment)
Assume the environment is (additively or multiplicatively) separable.
\end{assumption}

\begin{assumption}\label{ass:weights} (Concave Values) Let $A_i$, $B_i$ and $C_i$ be as defined in Definition \ref{def:separable}. Assume
  $A_i$ is differentiable and non-decreasing with respect to $\theta$
  in both cases (additive and multiplicative); in the additive case,
  $A_i$ is concave; in the multiplicative case, $A_i$ is log-concave;
  $B_i$ and $C_i$ are both bounded and non-negative.
\end{assumption}

\begin{assumption} ~\label{ass:mhr} (Monotone Hazard Rate)
The density of $F_i(\cdot)$ exists for every agent $i$ and is denoted
by $f_i$.
Also, the inverse hazard rate $\frac{1-F(\theta)}{f(\theta)}$ is decreasing in $\theta$.
\end{assumption}

\begin{theorem}\label{thm:main}
(Optimality) Suppose Assumptions \ref{ass:seperable},
\ref{ass:weights}  and \ref{ass:mhr} hold. Then, the Virtual Index
Mechanism (as defined in Figure 1) is optimal.
 Furthermore, the Virtual Index Mechanism is periodic ex-post incentive compatible and individually rational.
\end{theorem}

\begin{figure} ~\label{fig:mechanism}
\begin{center}
\fbox{\parbox{6in}{ \vspace*{2mm}

\i{\bf The Virtual Index Mechanism:}\\
\\
\i At (fictitious) time $t=0$,
\begin{description}
    \item\i\i	Each agent $i$ reports $\hat\theta_{i,0}$.
    \item\i\i	Each agent $i$ is charged $p_{i,0}$, see~Eq.\ (\ref{eq:piz}).
\end{description}
\i At each time $t = 1, \ldots$
\begin{description}
\item\i\i	Each agent $i$ reports $\thetaR_{i,t}$ and $\experienceRit$.
\item\i\i	Allocate to  $i^{\star}$, an agent with the maximum  $\GI_{i}^{\thetaRiz}(\thetaRit, \experienceRit, \infoit)$,
  see~Eq.\ (\ref{eq:virtual_index}).
\item\i\i     Charge $i^{\star}$, $p_{i^{\star},t}$, see~Eq.\ (\ref{eq:pit}).
\end{description}
} }
\end{center}
\caption{Description of the Virtual Index Mechanism}
\end{figure}

This theorem has a very surprising implication. The Virtual Index Mechanism is a mechanism designed to maximize revenue
in a context where the mechanism has to create appropriate incentives
for the agents to reveal private information over time.
However, the revenue it produces is \emph{identical} to the mechanism
that can (publicly) observe the dynamically evolving private
experiences of the agents (since it is also a feasible mechanism and optimal for the
problem with complete dynamic monitoring, and  by construction, it
has the same revenue).
Hence, the mechanism's capability to monitor the agents' dynamically
evolving private experiences does not yield any revenue for the
mechanism. We now formalize this claim.

\begin{corollary} Suppose Assumptions \ref{ass:seperable},
  \ref{ass:weights} and \ref{ass:mhr} hold. The Virtual Index
  Mechanism is optimal for the setting with complete dynamic
  monitoring. Furthermore, the seller obtains the same expected revenue in both the
  setting with complete dynamic monitoring and without
  monitoring.\end{corollary}

We now proceed to describe the Virtual Index Mechanism in detail. The
analysis of this theorem is contained in Subsection
\ref{sec:analysis}.

\subsection{Virtual Surplus,  Social Welfare, and a Fictitious Phase}

If agents are truthful, we seek that the allocation rule maximizes
the discounted sum of virtual values
$\virtuali(\theta_i,\experienceit,\infoit)$, as defined in Eq.\
(\ref{eq:virtual_value}). However, in order to satisfy incentive
constraints, we consider a broader mechanism, which allows the agents
to bid at a fictitious $t=0$ phase --- crucially, while we include
this fictitious $0$-th phase, we should point out that this phase
actually occurs at $t=1$.

Let us first understand the nature of the virtual surplus, under
separable values. A key observation is that separability of the value
functions implies that the virtual value
$\virtuali$ is an affine function of the
value, \emph{conditioned} on $\theta_i$.  Precisely,

\begin{lemma} (Affine Virtual Values) \label{lem:alapha_beta} For a separable value function, the virtual value is given by:
\begin{equation} \label{eq:virtual_value_separable}
\virtuali(\theta_i,\experienceit,\infoit) = \alpha_i(\theta_i) v_i(\theta_i, \experienceit,\infoit)  + \beta(\theta_i, \infoit)
\end{equation}
where functions $\alpha_i(\theta_i)$ and $\beta_i$ are defined as follows:
\begin{itemize}
\item (Additive) $\valuei(\theta_i,\experience_i,\info_i) = A_i(\theta_i,\info_i)+B_i(\experience_i,\info_i)$
\begin{eqnarray*}
\alpha_i(\theta_i)  & =  & 1 \\ 
\beta_i(\theta_i, \info_i)  & = & - \frac{1-F(\theta_i)}{f(\theta_i)} \frac{\partial A_i (\theta_i, \infoi)}{\partial \theta_i};
\end{eqnarray*}
\item (Multiplicative) $\valuei(\theta_i,\experience_i,\info_i) = A_i(\theta_i)B_i(\experience_i,\info_i) -C_i(\info_i)$
\begin{eqnarray*}
\alpha_i(\theta_i) & = &  1-\frac{1-F(\theta_i)}{f(\theta_i)}\frac{A^\prime_i(\theta_i)}{A_i(\theta_i)} \\ 
\beta_i(\theta_i, \info_i) & = & -\frac{1-F(\theta_i)}{f(\theta_i)} \frac{A^\prime_i(\theta_i)}{A_i(\theta_i)} C_i(\info_i).
\end{eqnarray*}
\end{itemize}
\end{lemma}

The observation is that once $\theta_i$ is fixed, the virtual values
are an affine function of the values --- though this affine function
could vary with time (in the additive $\beta(\cdot)$).  Recall, in a
static setting, affine transformations of the social welfare function
can be implemented via an affinely transformed VCG mechanism. Here, we
provide a (time varying) transformation for the dynamic setting, using an affine
transformation of the mechanism of \cite{BergemannV07}. This reduction
satisfies the dynamic incentive constraints and leaves us only with
the static incentive constraint required for eliciting initial types
$\theta$ --- this is the heart of our technical argument in Lemma~\ref{lem:theta_IC}.

The above structure motivates the use of a ``fictitious phase'', where we
break the $t=1$ phase into two parts. In the first part of this
phase (which we label as $t=0$), the agents make a report $r$ of
$\theta$, which we use to specify the functions $\alpha$ and
$\beta$. From then on, we proceed as an (affinely) transformed VCG
mechanism, where agents are allowed to re-report $\theta$ (though
$\alpha$ and $\beta$ are pegged to the initial report $r$). We now
specify this precisely.

\subsection{Allocation Rule}
To find the optimal allocation, we construct the following $(k+1)$-armed bandit process (where $k$ is the
number agents), augmented with a $0$-th arm which corresponds to a
non-transitioning arm that always pays $0$.
For a vector $r\in\R^k_{+}$,
define the \emph{weighted social welfare} as:
\begin{eqnarray}\label{eq:weighted_surplus}
 W^r(\theta, \experience,\info) =
\Expct\biggl[\sumt \deltat \sum_{i=1}^k q_{i,t} \Bigl( \alpha_i(r_i) v_i(\theta_i, \experienceit,\infoit)
+ \beta(r_i,\infoit)\Bigl) \biggl\vert \theta, \experience_1 = \experience, \info_1 = \info \biggl],
\end{eqnarray}
where for $i=0$, $v_0(\cdot)=\alpha_0(\cdot)=\beta_0(\cdot,\cdot)=0$.  Note
that the vector $r$ substitutes $\theta$ in the $\alpha$ and $\beta$
components of the virtual value (cf.\
Eq. (\ref{eq:virtual_value_separable})). With this structure, for $r =
\theta$, $W^\theta(\theta,e,\info)$ represents the virtual surplus and
the desired revenue of the mechanism (see\ Corollary
\ref{thm:upperbound}).

We use the initial phase $t=0$ to allow the agents to set
$r$. Subsequently, the allocation rule we use is the one that
maximizes the weighted social welfare for this given $r$. In particular,
for any $r$, we can find the algorithm that maximizes the weighted
social welfare using the Gittins index (see~\cite{Gittins89,
  Whittle82}).

\begin{definition} [Virtual index]
For each agent $i$, the virtual index is defined as:
%
\begin{eqnarray} \label{eq:virtual_index}
\GI_{i}^{r_i}(\theta_i, \experience_i, \info_i) =  \max_{\tau_i} \Expct\Biggl[\frac{\sum_{t=1}^{\tau_i} \deltat \xi^{r_i}(\theta_i, \experience_{it},\info_{it})}{\sum_{t=1}^{\tau_i} \deltat}
  \Biggl\vert \theta_i, \experience_{i1} = \experience_i, \info_{i1} = \info_i  \Biggl]
\end{eqnarray}

where the maximum is taken over all stopping times $\tau_i$ and
\begin{equation} \label{eq:xi}
\xi^{r_i}(\theta_i, \experience_{it},\info_{it}) = \alpha_i(r_i) v_i(\theta_i, \experience_{it},\info_{it}) +\beta(r_i,\info_{it}).
\end{equation}

\end{definition}

It is well-known that the \emph{Gittins index} policy (that which
chooses the arm with highest index) is the
(Bayes) optimal algorithm for multi-armed bandit problems. Hence, the allocation rule which chooses the highest virtual index is the optimal algorithm for the
$(k+1)$-armed bandit process where the goal is to maximize the weighted
social welfare (where $r$ is some fixed vector).

We now describe the allocation (specified in Figure~\ref{fig:mechanism}), including
reports and allocations. At time $t=0$, each agent reports $\thetaRiz$
their initial type $\theta_i$.  This initial report is used to
set $r$ above --- in other words, their initial report determines the
weights of the weighted social welfare function which the subsequent
allocation tries to maximize.  At ever subsequent time $t\geq 1$,
they report $\thetaRit$ and $\experienceRit$ (the component $\infoit$
is observed directly to the mechanism). This ``freedom to correct"
earlier misreports of $\theta_i$ leads to the stronger (periodic ex-post) notion of
incentive compatibility.

\subsection{Payment Rule}
We now construct a payment scheme that makes the mechanism periodic ex-post incentive compatible.
It turns out that it is only the $t=0$ fictitious phase where all
agents could potentially make a payment --- thereafter, only the agent
who is allocated pays. While, as specified in Figure 1, these payments
occur before the allocation, this is inconsequential (as they could
occur after the first allocation with no change to any of our guarantees).

As mentioned, the mechanism is the reminiscent of a static (affine) VCG mechanism, but with the added dynamic
twist of a time-varying, additive offset. In
~\cite{CavalloPS06,BergemannV07}, the payment of an agent after the
allocation corresponds should be equal to the {\em externality} he imposes to
other agents. In our context, the payment is an affine transformation of the externality imposed.  In particular, if at time $t$ the item is allocated to agent $i$, $i$
pays the following amount:
\begin{equation}\label{eq:pit}
p_{i,t} = \frac{(1-\delta)W^{\thetaR_0}_{-i,t}(\thetaR_t, \experienceRt, \infot) - \beta_i(\thetaRiz,\infoit)}{\alpha_i(\thetaRiz)},
\end{equation}
where $W_{-i,t}^r$ is the {\em optimal virtual surplus} of the other agents. Namely
\begin{eqnarray*}
 W^r_{-i,t}(\theta, \experience,\info) =
 \max_{\calQ \in \mathbb{Q}}
\Expct\left[\sum_{t'=t}^\infty \sum_{j\neq i}  \delta^{t'-t}
q_{j,t'} \xi^{r_j}(\theta_j, \experience_{j,t'},\info_{j,t'})
\middle \vert \theta,
  \experience_t = \experience, \infot = \info  \right]
\end{eqnarray*}
where $\mathbb{Q}$ is the set of allocations rules (and $j$ is summing over
the other $k$ arms, including the non-paying arm of $0$). Also $\xi$ is defined in Eq.~(\ref{eq:xi}).

Finally, we specify the payment at time $0$, $p_{i,0}$.  First define
$\gPricei(\thetaR)$ as:
%
\begin{eqnarray} \label{eq:ptheta}
\gPricei(\thetaR) = 
\gValuei(\thetaR)
 -  \int_0^{\thetaRi}
\Expct\left[\sumt \deltat \gallocit
\frac{\partial \valuei(z,\experienceit,\infoit)}{\partial z}
\middle\vert \thetaRi=z,\thetaRnoti \right] dz
\end{eqnarray}

Note this is the desired payment of agent $i$ conditioned on the
vector $\thetaRz$ of reports of initial type at times $0$ --- this revenue maximizes the
upper bound in Corollary~\ref{thm:upperbound}. The price charged
(to each agent $i$) is $\gPricei(\thetaRz)$ with a negative term
to offset all expected future payments:
\begin{equation}\label{eq:piz}
p_{i,0} = \gPricei(\thetaRz) - \Expct\left[\sumt \deltat q_{i,t} p_{i,t}  \middle\vert \hat{\theta}_0 \right]
\end{equation}
This offset allows, in expectation, the revenue to be
$\gPricei(\thetaRz)$ as desired. The heart of the proof is verifying incentive
compatibility with respect to the initial report $r$.

\subsection{Analysis of Theorem \ref{thm:main}} \label{sec:analysis}

We now give the outline of the proof of Theorem~\ref{thm:main} using a series of lemmas.
The proofs of these lemmas are given in the Appendix~\ref{sec:proofs}. In the discussion of this subsection, we focus on the issue of incentive compatibility, and address the issue of individual rationality in the proof of the lemmas.

The first step of the proof is to show that the mechanism is periodic ex-post incentive compatible for periods $t \geq 1$, irrespective of the reports of the initial types $\hat{\theta}_0$ at the (fictitious) period 0. Recall that the mechanism implements an efficient allocation with respect to the ``weights" that are
assigned as a function of the initial reports; the proof technique is similar to~\cite{BergemannV07}.

\begin{lemma}\label{lemma:vcg}
Let Assumption~\ref{ass:seperable} hold. Then, for any initial report $\hat{\theta}_0$,
the Virtual Index Mechanism is periodic ex-post incentive compatible and individually rational for periods $t\geq 1$.
\end{lemma}

The lemma above guarantees that, under the Virtual Index Mechanism, it is always a best response for agents to report their types truthfully regardless of the history, at any time $t \geq 1$ (assuming that other agents will be truthful in the future).

Therefore, we need only concern ourselves with period 0 deviations from the truthful strategy. To obtain incentive compatibility at time 0, we need some notion of monotonicity of the \emph{future allocations} with respect to period 0's report. The following lemma provides this monotonicity result.

Before we show the monotonicity lemma, we introduce some notation. Denote the Virtual Index (immediate) allocation rule by $q^{\rinit}(\theta,\experience,\info)$, where $\rinit$ is the initial report. That is,
for each $\rinit$, we have a (subsequent) allocation rule $q^{\rinit}(\cdot)$  which, at any time $t\geq 1$ assigns the item based on the
reported state $(\hat\theta_t,\hat\experience_t,\info_t)$.

\begin{lemma}~\label{lemma:monotone_q}
(Monotonic Allocation) Let Assumptions~\ref{ass:seperable},~\ref{ass:weights} and~\ref{ass:mhr} hold.
Then, for all (joint) states $(\theta,\experience,\info)$ and any two initial reports $\rinit$  and $\rinit^\prime$,
which only differ in the $i$-th coordinate and $\rinit_i\geq \rinit_i^\prime$, we have for the Virtual Index Mechanism that
\[
q_i^{\rinit}(\theta,\experience,\info) \geq q_i^{\rinit^\prime}(\theta,\experience,\info).
\]
\end{lemma}

Note that the lemma above defines an instantaneous notion of monotonicity: it establishes that at any time $t \geq 1$ and any reported state $(\hat\theta_t,\hat\experience_t,\info_t)$, the allocation of the item to agent $i$ is more likely if his first period report $\hat{\theta}_{i,0}$ is higher. The underlying multi-armed bandit stochastic process guarantees that this notion of monotonicity is sufficient for agent $i$'s expected discounted sum of all his future values to be monotonic on $\hat{\theta}_{i,0}$. We, therefore, obtain in the following lemma that the Virtual Index Mechanism is also incentive compatible at period 0. This lemma is the key component of our technical argument.

\begin{lemma} \label{lem:theta_IC}
Under Assumptions~\ref{ass:seperable},~\ref{ass:weights} and~\ref{ass:mhr},
the Virtual Index Mechanism is both periodic ex-post incentive compatible
and individually rational.
\end{lemma}

We can now state the proof of our main theorem.

\begin{proof}[{\bf Proof of  Theorem \ref{thm:main}}]
From Lemma \ref{lem:theta_IC}, we obtain that the Virtual Index Mechanism is both periodic ex-post incentive compatible
and individually rational.
Hence, From Eq.\ (\ref{eq:piz}), we get that the revenue produced by the Virtual Index Mechanism is equal to
\[\rev = \sum_{i=1}^k \sum_{t=0}^\infty \Expct[p_{i,t}] = \sum_{i=1}^k P_i(\hat{\theta}_0),\] where $P_i(\cdot)$ is as defined in Eq.\ (\ref{eq:ptheta}).
 This value is equal to the bound given in Corollary \ref{thm:upperbound}. Therefore, the mechanism is optimal.
\end{proof}

\medskip

\bibliographystyle{plain}
\bibliography{mechanismdesign}

\begin{thebibliography}{21}
\providecommand{\natexlab}[1]{#1}
\providecommand{\url}[1]{\texttt{#1}}
\expandafter\ifx\csname urlstyle\endcsname\relax
  \providecommand{\doi}[1]{doi: #1}\else
  \providecommand{\doi}{doi: \begingroup \urlstyle{rm}\Url}\fi

\bibitem[Agarwal et~al.(2009)Agarwal, Athey, and Yang]{AgrawalAY09}
Nikhil Agarwal, Susan Athey, and David Yang.
\newblock Skewed bidding in pay per action auctions for online advertising.
\newblock \emph{American Economic Review Papers and Proceedings}, 2009.

\bibitem[Athey and Segal(2007)]{AtheyS07}
Susan Athey and Ilya Segal.
\newblock An efficient dynamic mechanism.
\newblock \emph{Working paper}, 2007.

\bibitem[Babaioff et~al.(2009)Babaioff, Sharma, and Slivkins]{BabaioffSS09}
Moshe Babaioff, Yogeshwer Sharma, and Aleksandrs Slivkins.
\newblock Characterizing truthful multi-armed bandit mechanisms.
\newblock In \emph{ACM Conference on Electronic Commerce}, pages 79--88, 2009.

\bibitem[Battaglini(2005)]{Battaglini05}
Marco Battaglini.
\newblock Long-term contracting with markovian customers.
\newblock \emph{American Economic Review}, 95\penalty0 (3):\penalty0 637--658,
  2005.

\bibitem[Bergemann and V\"alim\"aki(2007)]{BergemannV07}
Dirk Bergemann and Juuso V\"alim\"aki.
\newblock The dynamic pivot mechanism.
\newblock \emph{Working paper}, 2007.

\bibitem[Cavallo et~al.(2006)Cavallo, Parkes, and Singh]{CavalloPS06}
Ruggiero Cavallo, David~C. Parkes, and Satinder~P. Singh.
\newblock Optimal coordinated planning amongst self-interested agents with
  private state.
\newblock In \emph{Proceedings of the 22nd Conference in Uncertainty in
  Artificial Intelligence}, 2006.

\bibitem[Cavallo et~al.(2007)Cavallo, Parkes, and Singh]{CavalloPS07}
Ruggiero Cavallo, David~C. Parkes, and Satinder Singh.
\newblock Efficient online mechanisms for persistent, periodically inaccessible
  self-interested agents.
\newblock \emph{Working paper}, 2007.

\bibitem[Deb(2008)]{Deb08}
Rahul Deb.
\newblock Optimal contracting of new experience goods.
\newblock \emph{Working paper}, 2008.

\bibitem[\"Eso and Szentes(2007)]{EsoS07}
Peter \"Eso and Balazs Szentes.
\newblock Optimal information disclosure in auctions and the handicap auction.
\newblock \emph{Review of Economic Studies}, 74\penalty0 (3):\penalty0
  705--731, 2007.

\bibitem[Gittins(1989)]{Gittins89}
J.~C. Gittins.
\newblock \emph{Allocation Indices for Multi-Armed Bandits}.
\newblock Wiley, London, 1989.

\bibitem[Mahdian and Tomak(2007)]{MahdianT07}
Mohammad Mahdian and Kerem Tomak.
\newblock Pay-per-action model for online advertising.
\newblock In \emph{Internet and Network Economics, Third International
  Workshop}, pages 549--557, 2007.

\bibitem[Milgrom and Segal(2002)]{MilgromS02}
Paul Milgrom and Ilya Segal.
\newblock Envelope theorems for arbitrary choice sets.
\newblock \emph{Econometrica}, 70\penalty0 (2):\penalty0 583--601, 2002.

\bibitem[Myerson(1981)]{Myerson81}
Roger Myerson.
\newblock Optimal auction design.
\newblock \emph{Mathematics of Operations Research}, 1981.

\bibitem[Myerson(1986)]{Myerson86}
Roger Myerson.
\newblock Multistage games with communications.
\newblock \emph{Econometrica}, 54\penalty0 (2):\penalty0 323--358, 1986.

\bibitem[Nazerzadeh et~al.(2008)Nazerzadeh, Saberi, and Vohra]{NazerzadehSV08}
Hamid Nazerzadeh, Amin Saberi, and Rakesh Vohra.
\newblock Dynamic cost-per-action mechanisms and applications to online
  advertising.
\newblock In \emph{Proceedings of the 17th International Conference on World
  Wide Web}, pages 179--188, 2008.

\bibitem[Pai and Vohra(2008)]{PaiV08}
Mallesh Pai and Rakesh Vohra.
\newblock Optimal dynamic auctions.
\newblock \emph{Working paper}, 2008.

\bibitem[Parkes and Singh(2003)]{ParkesS03}
David~C. Parkes and Satinder~P. Singh.
\newblock An mdp-based approach for online mechanism design.
\newblock In \emph{Proceedings of the 17th Conference on Neural Information
  Processing Systems}, 2003.

\bibitem[Pavan et~al.(2008)Pavan, Segal, and Toikka]{PavanST08}
Alessandro Pavan, Ilya Segal, and Juuso Toikka.
\newblock Dynamic mechanism design: Incentive compatibility, profit
  maximization and information disclosure.
\newblock \emph{Working paper}, 2008.

\bibitem[Pavan et~al.(2009)Pavan, Segal, and Toikka]{PavanST09}
Alessandro Pavan, Ilya Segal, and Juuso Toikka.
\newblock Infinite-horizon mechanism design: The independent-schock approach.
\newblock \emph{Preliminary Draft}, 2009.

\bibitem[Vulcano et~al.(2002)Vulcano, van Ryzin, and Maglaras]{VulcanoRM02}
Gustavo Vulcano, Garrett van Ryzin, and Costis Maglaras.
\newblock Optimal dynamic auctions for revenue management.
\newblock \emph{Management Science}, 48\penalty0 (11):\penalty0 1388--1407,
  2002.

\bibitem[Whittle(1982)]{Whittle82}
Peter Whittle.
\newblock \emph{Optimization Over Time}, volume~1.
\newblock Wiley, Chichester, 1982.

\end{thebibliography}


\appendix

\section{Appendix}

\subsection{Analysis of Theorem \ref{thm:ICmonitoring}} \label{sec:envelope}

The proof of this theorem follows the outline of \cite{Myerson81} for establishing incentive compatibility in static mechanism design. In Lemma \ref{lem:env}, we derive an envelope condition that the utility of players participating in incentive compatible mechanism must satisfy. We then use this envelope condition to derive the desired result.

\begin{lemma}\label{lem:env}
Assume complete dynamic monitoring. Assume as well that the partial derivative $\frac{\partial  \valuei(\theta_i,\experienceit,\infoit)}{\partial \theta_i}$ exists for all  $\theta_i$, $\experienceit$ and $\infoit$ and there exists some $B < \infty$ such that $|\frac{\partial  \valuei(\theta_i,\experienceit,\infoit)}{\partial \theta_i}| \leq B$ for all  $\theta_i$, $\experienceit$ and $\infoit$.  Then, any incentive compatible mechanism satisfies for all $i$, $\theta_i$ and $\theta_{-i}$, for all players other $i$ being truthful,
\begin{equation} \label{eq:env}
U_i^\calM(\theta_i)- U_i^\calM(0,\theta_{-i}) = \int_0^{\theta_i}
\Expct\left[\sumt \deltat \allocit
\frac{\partial \valuei(z,\experienceit,\infoit)}{\partial z}
\middle\vert \theta_i=z, \theta_{-i}\right] dz.
\end{equation}
\end{lemma}
\begin{proof}
In a direct mechanism in this setting, the agents' only report is $\hat{\theta}$ at the initial period. The mechanism design problem is therefore static and we use the classical results from \cite{MilgromS02} (Theorem 2) to obtain our envelope condition. We now show that these conditions are satisfied here.

For any mechanism $\calM$ and initial type profile $\theta$, let the expected utility of player $i$ reporting type $\hat{\theta}_i$ is be given by
\begin{equation*}
\hat{U}_i^\calM(\hat{\theta}_i,\theta_{i}|\theta_{-i}) = \Expct\left[\sumt \deltat \left(\allocit \valuei(\theta_i,\experienceit,\infoit) - p_{i,t}(\hat{\theta},\experienceit,\infoit) \right)\middle\vert \theta_{-i}\right].
\end{equation*}
Consider the term $\hat{U}_i^\calM(\hat{\theta}_i, \cdot |\theta_{-i})$ applied to two different values $\theta_i$ and $\theta_i'$. Taking the difference between the two and dividing by $\theta_i - \theta_i'$, we obtain
\begin{equation*}
\frac{\hat{U}_i^\calM(\hat{\theta}_i,\theta_{i}|\theta_{-i}) -\hat{U}_i^\calM(\hat{\theta}_i,\theta_{i}'|\theta_{-i})}{\theta_i - \theta_i'} = \Expct\left[\sumt \deltat \allocit \left(\frac{ \valuei(\theta_i,\experienceit,\infoit) - \valuei(\theta_i',\experienceit,\infoit)}{\theta_i-\theta_i'}\right)\middle\vert \theta_{-i}\right].
\end{equation*}
Since $|\allocit| \leq 1$ for all $\hat{\theta}$, $\experienceit$ and $\infoit$, and the partial derivative $\frac{ \partial \valuei(\theta_i,\experienceit,\infoit)}{\partial \theta_i}$ exists for all $\hat{\theta}$, $\experienceit$ and $\infoit$ and is bounded by $B$, we use Lebesgue's Dominated Convergence Theorem to obtain that the partial derivative $\frac{\partial \hat{U}_i^\calM(\hat{\theta}_i,\theta_{i}|\theta_{-i})}{\partial \theta_i}$ exists for all $\theta_i$ and $\hat{\theta}$ and satisfies
\begin{equation*}
\frac{\partial \hat{U}_i^\calM(\hat{\theta}_i,\theta_{i}|\theta_{-i})}{\partial \theta_i} = \Expct\left[\sumt \deltat \allocit \frac{ \partial \valuei(\theta_i,\experienceit,\infoit)}{\partial \theta_i} \middle\vert \theta_{-i}\right].
\end{equation*} Furthermore, $|\frac{\partial \hat{U}_i^\calM(\hat{\theta}_i,\theta_{i}|\theta_{-i})}{\partial \theta_i}| \leq \frac{B}{1-\delta}$ for all $\hat{\theta}_i$ and $\theta$ and, therefore, $\hat{U}_i^\calM(\hat{\theta}_i,\cdot|\theta_{-i})$ is absolutely continuous for any $\hat{\theta}_i$ and $\theta_{-i}$. Therefore, the function $\hat{U}_i^\calM(\hat{\theta}_i,\theta_i|\theta_{-i})$ satisfies the conditions of \cite{MilgromS02} Theorem 2, yielding Eq.\ (\ref{eq:env}).
\end{proof}

\begin{proof}[{\bf Proof of Theorem \ref{thm:ICmonitoring}}]
 Consider first the utility $U_i^\calM(\theta)$ of an agent $i$ under an initial type profile $\theta$, which is given by
\begin{equation*}U_i^\calM(\theta) - U_i^\calM(0,\theta_{-i}) = \int_0^{\theta_i} \Expct\left[\sumt \deltat \allocit\frac{\partial \valuei(z,\experienceit,\infoit)}{\partial z} \middle\vert \theta_i=z, \theta_{-i}\right] dz\end{equation*} from Eq.\ (\ref{eq:env}). Taking the expectation of this term over all possible type profiles $\theta$, we obtain
\begin{equation*}\Expct[U_i^\calM(\theta) - U_i^\calM(0,\theta_{-i}) ] = \int_0^{\Theta_i} \int_0^{\theta_i} \Expct\left[\sumt \deltat \allocit\frac{\partial \valuei(z,\experienceit,\infoit)}{\partial z} \middle\vert \theta_i=z\right] dz f_i(\theta_i) d\theta_i.\end{equation*} Inverting the order of integration,
\begin{eqnarray*}\Expct[U_i^\calM(\theta) - U_i^\calM(0,\theta_{-i}) ] &=& \int_0^{\Theta_i} \int_z^{\Theta_i} \Expct\left[\sumt \deltat \allocit\frac{\partial \valuei(z,\experienceit,\infoit)}{\partial z} \middle\vert \theta_i=z\right] f_i(\theta_i) d\theta_i dz\\ &=&
\int_0^{\Theta_i} \Expct\left[\sumt \deltat \allocit\frac{\partial \valuei(z,\experienceit,\infoit)}{\partial z} \middle\vert \theta_i=z\right] (1-F_i(z)) dz.\end{eqnarray*} By multiplying and dividing the right-hand side of the equation above by the density $f_i(z)$ we obtain an unconditional expectation,
\begin{equation}\label{eq:util env}\Expct[U_i^\calM(\theta) - U_i^\calM(0,\theta_{-i}) ] =
\Expct\left[\sumt \deltat \allocit \frac{1-F_i(\theta_i)}{f_i(\theta_i)}\frac{\partial \valuei(\theta_i,\experienceit,\infoit)}{\partial \theta_i}\right].\end{equation} Now note that the total revenue of the mechanism is given by the sum of the payments from the agents, which themselves are the difference between the value of the allocations to the agents and the utility they obtain, i.e.,
\[ \rev^\calM = \sum_{i=1}^k \Expct [P_i^{\calM} (\theta)]  = \Expct [V_i^{\calM}(\theta) - U_i^\calM(\theta)].\] Combining the definition of the value $V_i^{\calM}(\theta)$ with Eq.\ (\ref{eq:util env}),
\[ \rev^\calM = \sum_{i=1}^k \Expct \left[\sumt \deltat \allocit \valuei(\theta_i,\experienceit,\infoit) - \sumt \deltat \allocit \frac{1-F_i(\theta_i)}{f_i(\theta_i)}\frac{\partial \valuei(\theta_i,\experienceit,\infoit)}{\partial \theta_i} - U_i^\calM(0,\theta_{-i})\right],
\] yields the desired result by the plugging in the definition of the virtual value $\virtual$.
\end{proof}

\newpage

\subsection{Omitted Proofs} ~\label{sec:proofs}

\begin{proof} [{\bf Proof of Lemma~\ref{lem:alapha_beta}}]
The claim for additive separability is trivial.
In the multiplicatively separable case, we have:
\begin{eqnarray*}
\virtuali(\theta_i,\experience_i,\info_i)  &= &
A_i(\theta_i)B_i(\experience_i,\info_i) -C_i(\info_i)-
\frac{1-F_i(\theta_i)}{f_i(\theta_i)} A_i^\prime(\theta_i,\info_i)
B_i(\experience_i,\info_i) \\
&=&\left(A_i(\theta_i)-\frac{1-F_i(\theta_i)}{f_i(\theta_i)} A_i^\prime(\theta_i,\info_i)\right)
B_i(\experience_i,\info_i) -C_i(\info_i)
 \\
&=&\alpha_i(\theta_i) A_i(\theta_i) B_i(\experience_i,\info_i) -C_i(\info_i)
 \\
&=&\alpha_i(\theta_i) \valuei(\theta_i,\experience_i,\info_i) +(\alpha_i(\theta_i)-1)C_i(\info_i)\\
&=&\alpha_i(\theta_i) \valuei(\theta_i,\experience_i,\info_i) +\beta_i(\theta_i,\info_i)
\end{eqnarray*}
and the claim follows.
\end{proof}

\medskip

\begin{proof}[{\bf Proof of Lemma~\ref{lemma:vcg}}]
The proof follows the outline of \cite{BergemannV07}, with ``virtual values" replacing values as the mechanism's objective. In this proof we assume an initial type report profile $\hat{\theta}_0$ (not necessarily truthful reports) and concern ourselves only with periods $t \geq 1$.
Let $W^{\calR_i}_{t}(\theta, \experience,\info)$ denote the discounted future virtual surplus, at time $t\ge1$ and state $(\theta, \experience,\info)$,
when agent $i$ uses reporting strategy $\calR_i$ and other agents are truthful.
\begin{eqnarray*}
W^{\calR_i}_{t}(\theta, \experience,\info) & = &
 \max_{\calQ \in \mathbb{Q}} \Expct\left[\sum_{t'=t}^\infty \sumi  \delta^{t'-t} q^{\calR_i}_{j,t'}
 \left(\alpha_j(\thetaR_{j,0}) v_j(\theta_j, \experience_{j,t'},\info_{j,t'})  +  \beta(\thetaR_{j,0},\info_{j,t'}) \right)
\middle \vert \theta,
  \experience_t = \experience, \infot = \info  \right]
\end{eqnarray*}
where $\mathbb{Q}$ is the set of all allocation rules and $q^{\calR_i}_{j,t'}$ denotes the allocation induced by $\calR_i$
(other agents are assumed truthful).

Similarly define virtual surplus without agent $i$:
\begin{eqnarray*}
W_{-i,t}(\theta, \experience,\info) & = &
 \max_{\calQ \in \mathbb{Q}_{-i}} \Expct\left[\sum_{t'=t}^\infty \sum_{j\neq i}  \delta^{t'-t} q_{j,t'}
 \left(\alpha_j(\thetaR_{j,0}) v_j(\theta_j, \experience_{j,t'},\info_{j,t'})  +  \beta(\thetaR_{j,0},\info_{j,t'}) \right)
\middle \vert \theta,
  \experience_t = \experience, \infot = \info  \right],
\end{eqnarray*}
where $\mathbb{Q}_{-i}$ is the set of allocation rules that never assign items to agent $i$.  Note $W_{-i}$ is defined without reference to ${\calR_i}$ because agent $i$ does not effect the allocation.
In addition, let $m_{i,t}$ denote the marginal contribution of agent $i$, at time $t$, to the virtual surplus, i.e.,
\begin{equation} \label{eq:mi}
m_{i,t} = W^{\calR_i}(\theta, \experiencet,\infot) - W_{-i}(\theta, \experiencet,\infot) - \delta \Expct[W^{\calR_i}(\theta, \experience_{t+1},\infotplus) - W_{-i}(\theta, \experience_{t+1},\infotplus)].
\end{equation}

If the mechanism does not allocate the item to agent $i$ at time $t$, then $m_{i,t} = p_{i,t} = 0$.
But, if the mechanism does allocate the item to agent $i$ at time $t$, then
$$ W^{\calR_i}(\theta, \experiencet,\infot) =  \left(\alpha_i(\thetaRiz) v_i(\theta_i, \experienceit, \infoit) + \beta(\thetaRiz, \infoit)\right)  + \delta \Expct[W^{\calR_i}(\theta, \experience_{t+1},\infotplus)].$$
Since an agent's state does not change in the absence of an allocation, if the item is allocated to $i$ at time $t$, we have
$$  W_{-i}(\theta, \experiencet,\infot) = W_{-i}(\theta, \experience_{t+1},\infotplus).$$
Hence, we obtain a marginal contribution (cf.\ Eq.~(\ref{eq:mi})) for agent $i$ at time $t$ of
\begin{eqnarray*}
m_{i,t}  & = &  \left(\alpha_i(\thetaRiz) v_i(\theta_i, \experienceit, \infoit) + \beta(\thetaRiz, \infoit)\right)   -  (1-\delta)  W_{-i}(\statet)  \\
& = & \alpha_i(\thetaRiz) \left(v_i(\theta_i, \experienceit, \infoit) -  p_{i,t}\right),
\end{eqnarray*} where the price $p_{i,t}$ is given in Eq.\ (\ref{eq:pit}). Noting that $m_{i,t}$ can only be different than zero if $q^{\calR_i}_{i,t}=1$, we obtain that the expected future utility of agent $i$ at time $t$ given reporting strategy $\calR_{i,t}$ is
\begin{eqnarray*}
U^{\calR_i}_{i,t}(\theta, \experienceit, \infoit) & = & \Expct\left[\sum_{t'=t}^{\infty} \delta^{t'-1} (q^{\calR_i}_{i,t'}  v_i(\theta_i, \experienceit, \infoit) -  p_{i,t})\right] \\
 & = & {1\over\alpha_i(\thetaRiz)} \Expct\left[\sum_{t'=t}^{\infty} \delta^{t'-t}  m_{i,t}\right] \\
 & = & {1\over\alpha_i(\thetaRiz)} \left(W^{\calR_i}(\theta, \experiencet, \infot) - W_{-i}(\theta, \experiencet, \infot)\right).
\end{eqnarray*}

Note that $W_{-i}$ is independent of all of agent $i$'s reports.
Also, $W^{\calR_i}_{-i}(\thetaRt, \experienceRt,\infot)$ is maximized if $i$ reports truthfully
since $W$ is define as the maximum virtual surplus obtained by an allocation with respect to the true (joint) state. Therefore, we obtain that the
mechanism  is periodic ex-post incentive compatible. Observe as well that $W^{\calT} \ge W_{-i}$, where $\calT$ denotes the truthful strategy; yielding that the mechanism is also periodic ex-post individually rational.

\end{proof}

\medskip

\begin{proof}[{\bf Proof of Lemma~\ref{lemma:monotone_q}}] The Virtual Index Mechanism allocates the item to the agent with the highest ``virtual index". Therefore, it is sufficient to show that the virtual value $\virtuali$ is non-decreasing on the initial report of the agent. Therefore, it suffices to show that
both $\alpha_i(\cdot)$ and $\beta_i(\cdot,\info)$ are non-decreasing functions of
$\theta_i$ under the assumptions of the lemma. With this it directly
follows that the Virtual Index is monotonic in the initial reports.
Let $\eta_i(\theta_i)$ denote the hazard rate, i.e.,
$$\eta_i(\theta_i) = \frac{f_i(\theta_i)}{1-F_i(\theta_i)}.$$
In the additive case,
\[
\beta_i^\prime(\theta_i,\info_i) = \frac{\eta_i^\prime(\theta_i)}{\eta_i^2(\theta_i)} A_i^\prime(\theta_i) -\frac{1}{\eta_i(\theta_i)}A^{\prime\prime}(\theta_i)
\]
where $(\cdot)^\prime$ denotes a partial derivative with respect to $\theta_i$. By the assumptions that $A_i$ is concave and non-decreasing,
 and the hazard rate is non-negative and increasing, we have that the above
is non-negative.

In the multiplicative case, first note that $\alpha_i(\theta_i) =
1-\frac{1}{\eta_i(\theta_i)}(\log A_i(\theta_i))^\prime$, so
\[
\alpha_i^\prime(\theta_i) = \frac{\eta_i^\prime(\theta_i)}{\eta_i^2}\frac{A_i^\prime(\theta_i)}{A_i(\theta_i)} -
\frac{1}{\eta_i(\theta_i)} (\log A_i(\theta_i))^{\prime \prime}
\]
which is non-negative by the assumptions. Since $\beta_i =(\alpha_i-1)C_i$
 and since $C_i$ is positive,
$\beta_i$ is also non-decreasing.
\end{proof}

\medskip

\begin{proof}[{\bf Proof of Lemma~\ref{lem:theta_IC}}]

Let $U_i(\theta)$ be the expected utility of $i$ in under the Virtual
Index Mechanism conditioned on the initial types being $\theta$ (and
everyone behaving truthfully).
By construction, see Eq.~(\ref{eq:ptheta}), we have:
\begin{eqnarray} \label{eq:p1}
U_i(\theta)
 =  \gValuei(\theta) - \gPricei(\theta)
 =  \int_0^{\theta_i}
\Expct\left[\sumt \deltat \gallocit
\frac{\partial \valuei(z,\experienceit,\infoit)}{\partial z}
\middle\vert \theta_i=z,\theta_{-i} \right] dz
\end{eqnarray}
Here, we are slightly abusing
notation in that we are only specifying the superscript with respect to $z$
in $\gallocit$ (it implicitly also depends on $\theta_{-i}$).
Observer that $U_i(0)=0$. Moreover,
because $v_i$ is increasing in $\theta_i$, $U_i$ is non-negative.
Hence, the mechanism is ex-post individually rational with respect to the (first) report of the initial type.
We now prove that the mechanism is ex-post incentive compatible with respect to the (first) report of the initial type.

Let $\tilde U_i(\theta;\theta_i^\prime)$ be the utility of agent $i$
conditioned on: the initial types being $\theta$; the initial report
in the $0$-th period of $i$ being $\theta_i^\prime$; where $i$ behaves
optimally thereafter; and where all other agents behave truthfully
(conditioned on their initial type and report being $\theta_{-i}$). To
prove (periodic ex-post) incentive compatibility, we must show that for all $\theta$ and all
$\theta_i^\prime$,
\[
U_i(\theta) \geq \tilde U_i(\theta;\theta_i^\prime).
\]

Let $\theta^\prime$ equal $\theta$ in all coordinates except in
coordinate $i$, where it equals to $\theta_i^\prime$.
Lemma~\ref{lemma:vcg} shows that truthfulness is the
optimal continuation strategy from time $t\geq 1$ onwards.
Hence,
\[
U_i(\theta^\prime) = \tilde U_i(\theta^\prime;\theta_i^\prime).
\]
Thus, it suffices to show:
\begin{equation}\label{eq:suffices}
U_i(\theta)-U_i(\theta^\prime) \geq
\tilde U_i(\theta;\theta_i^\prime)-\tilde U_i(\theta^\prime;\theta_i^\prime)
\end{equation}
for all $\theta$ and $\theta_i^\prime$. Now we write this condition
more explicitly.

Now let us consider the mechanism $q^{\theta^\prime}(\cdot)$ from
$t\geq1$, where $\theta^\prime$ is fixed. Let
$U_i^{\theta^\prime}(\theta)$ be the utility of this mechanism from
$t\geq 1$ onwards (where truthful reporting occurs).  By Lemma~\ref{lemma:vcg}, we have that $q^{\theta^\prime}$ is an
incentive compatible allocation.  Hence, the envelope
lemma implies:\footnote{The envelope lemma requires that $|\frac{\partial v_i(\theta_i,e_i,\infoi)}{\partial \theta_i}| \leq B$ for some $B < \infty$ for all $\theta_i$, $e_i$ and $\info_i$. Assumption \ref{ass:weights}, together with the compact support of $\theta_i$, guarantee that this condition holds.}
\[
U_i^{\theta^\prime}(\theta)- U_i^{\theta^\prime}(\theta^\prime)
= \int_{\theta_i^\prime}^{\theta_i}
\Expct\left[\sumt \delta^t \allocit^{\theta^\prime}
\frac{\partial \valuei(z,\experienceit,\infoit)}{\partial z}
\middle\vert\theta_i=z,\theta_{-i}\right] dz
\]
Now note that at $t=0$ the charge only
depends on the initial report (e.g., $\theta^\prime$). Hence,
\begin{equation}  \label{eq:p2}
\tilde U_i(\theta;\theta_i^\prime)-\tilde U_i(\theta^\prime;\theta_i^\prime)
=U_i^{\theta^\prime}(\theta)- U_i^{\theta^\prime}(\theta^\prime)
= \int_{\theta_i^\prime}^{\theta_i}
\Expct\left[\sumt \deltat \allocit^{\theta^\prime}
\frac{\partial \valuei(z,\experienceit,\infoit)}{\partial z}
\middle\vert\theta_i=z,\theta_{-i}\right] dz
\end{equation}

where now the evolution of the process is under the mechanism
$q^{\theta^\prime}$.

Combining Eq~(\ref{eq:p1}) and~(\ref{eq:p2}), we have that Eq.~(\ref{eq:suffices}) is equivalent to:
\begin{align*}
\int_{\theta_i^\prime}^{\theta_i}
\Expct\left[\sumt \deltat \gallocit
\frac{\partial \valuei(z,\experienceit,\infoit)}{\partial z}
\middle\vert\theta_i=z,\theta_{-i}\right] dz
\geq
 \int_{\theta_i^\prime}^{\theta_i}
\Expct\left[\sumt \deltat \allocit^{\theta^\prime}
\frac{\partial \valuei(z,\experienceit,\infoit)}{\partial z}
\middle\vert\theta_i=z,\theta_{-i}\right] dz
\end{align*}
We can couple the two expectations by sampling the initial types and an
infinite sequence of experiences --- then each process is evolved on
this sample using the appropriate allocations. Thus, this condition is
equivalent to:
\begin{align*}
\int_{\theta_i^\prime}^{\theta_i}
\Expct\left[\sumt \deltat \left(
\gallocit
\frac{\partial \valuei(z,\experienceit,\infoit)}{\partial z}
-
\allocit^{\theta^\prime}
\frac{\partial \valuei(z,\experienceit^\prime,\infoit^\prime)}{\partial z}
\right)
\middle\vert\theta_i=z,\theta_{-i}\right] dz
\geq 0
\end{align*}
where $\experienceit^\prime$ and $\infoit^\prime$ denote the experiences under
the allocation $q^{\theta^\prime}$.

Now let $\tau_k$ be the time at which the $k$-th allocation to $i$ occurs
under the $\gallocit$ (assumed to be infinite if the $k$-th allocation
does not occur), and let $\tau_k^\prime$ be this time under
$\allocit^{\theta^\prime}$. By our coupled process, we have that
$\experience_{i,\tau_k^\prime}^\prime$ equals $\experience_{i,\tau_k}$ for
all $k$. This is because after $k$ allocations, both private
experiences have advanced precisely $k-1$ times. Similarly, the
same is true for the public experiences. Hence, the above condition is
equivalent to:
\[
\int_{\theta_i^\prime}^{\theta_i}
\Expct\left[\sumk \left(
\delta^{\tau_k}
-
\delta^{\tau_k^\prime}
\right)
\frac{\partial \valuei(z,\experience_{i,\tau_k},\info_{i,\tau_k})}
{\partial z}
\middle\vert\theta_i=z,\theta_{-i} \right] dz
\]
since the experiences at $\tau_k$ and $\tau_k^\prime$ are identical.

Without loss of generality, assume that $\theta > \theta'$.
To complete the proof of incentive compatibility, we show that $\tau_k^\prime \geq \tau_k$,
using the monotonicity of the allocation
(Lemma~\ref{lemma:monotone_q}). This implies that:
\[
q_i^{\theta^\prime}(z,\experience_t,\info_t) \leq q_i^{z}(z,\experience_t,\info_t) \, .
\]
We proceed inductively. For $k=1$, note that until the first
allocation occurs in either process, the state of $i$ is identical at
each timestep under both allocations --- this is because the Gittins
allocation is an index mechanism (so the states of the other agents
advance identically when $i$ is not present). Hence, monotonicity
implies that the first allocation under $q^{\theta^\prime}$ cannot
occur before that time under $\galloc$. Taking the inductive step,
assume that $\tau_{k-1}^\prime \geq \tau_{k-1}$. If
$\tau_{k-1}^\prime\geq \tau_k$, then we are done. If not, this implies
at time $\tau_{k-1}^\prime$, agent $i$ has been allocated
precisely $k-1$ times in both processes, and so she has an identical
state in either process. Hence, an identical argument to that of the
first period shows that the $k$-th allocation (if it occurs) under
$q^{\theta^\prime}$ cannot occur before that under $\galloc$,
which completes the proof.
\end{proof}

\end{document}